\documentclass[12pt,epsf]{article}

\input psfig.sty

\newcommand{\Msun}{M_{\odot}}
\newcommand{\OI}{O\,{\sc i}}
\newcommand{\OII}{O\,{\sc ii}}
\newcommand{\CaII}{Ca~{\sc ii}}
\newcommand{\FeII}{Fe~{\sc ii}}
\newcommand{\NiII}{Ni~{\sc ii}}

\def\gsim{\mathrel{\rlap{\lower 4pt \hbox{\hskip 1pt $\sim$}}\raise 1pt \hbox {$>$}}} \def\lsim{\mathrel{\rlap{\lower 4pt \hbox{\hskip 1pt $\sim$}}\raise 1pt \hbox {$<$}}}

\newenvironment{sciabstract}{%
\begin{quote} \bf}
{\end{quote}}

\newcounter{lastnote}

\title{
Asphericity in Supernova Explosions \\ 
from Late-Time Spectroscopy} 
\author{
Keiichi Maeda$^{1,2,3,\ast}$, 
Koji Kawabata$^{4}$, 
Paolo A. Mazzali$^{2,5,6}$, 
Masaomi Tanaka$^{7}$, \\
Stefano Valenti$^{8,9}$, 
Ken'ichi Nomoto$^{1,6,7}$, 
Takashi Hattori$^{10}$, 
Jinsong Deng$^{11}$, \\
Elena Pian$^{5}$, 
Stefan Taubenberger$^{2}$,
Masanori Iye$^{12}$, \\
Thomas Matheson$^{13}$, 
Alexei V. Filippenko$^{14}$, 
Kentaro Aoki$^{10}$, \\
George Kosugi$^{15}$, 
Youichi Ohyama$^{16}$,
Toshiyuki Sasaki$^{10}$, and 
Tadafumi Takata$^{17}$
\\
\normalsize{$^{1}$Institute for the Physics and Mathematics of the Universe (IPMU), 
University of Tokyo, }\\
\normalsize{Kashiwa-no-ha 5-1-5, Kashiwa City, Chiba 277-8582, Japan}\\
\normalsize{$^{2}$Max-Planck-Institut f\"ur Astrophysik, 
Karl-Schwarzschild-Stra{\ss}e 1, 85741 Garching, Germany}\\
\normalsize{$^{3}$Department of Earth Science and Astronomy,  College of Arts and Science, }\\
\normalsize{University of Tokyo, Tokyo 153-8902, Japan}\\
\normalsize{$^{4}$Hiroshima Astrophysical Science Center, Hiroshima University, 
Hiroshima 739-8526, Japan}\\ 
\normalsize{$^{5}$National Institute for Astrophysics-OATs, Via G.B.Tiepolo, 11, 
34143 Trieste, Italy}\\ 
\normalsize{$^{6}$Research Center for the Early Universe, School of Science, University of Tokyo, }\\
\normalsize{Tokyo 113-0033, Japan}\\
\normalsize{$^{7}$Department of Astronomy, School of Science,  University of Tokyo, }\\
\normalsize{Tokyo 113-0033, Japan,}\\
\normalsize{$^{8}$Physics Department, University of Ferrara, I-44100 Ferrara, Italy}\\
\normalsize{$^{9}$European Organization for Astronomical Research in the Southern Hemisphere, }\\
\normalsize{Karl-Schwarzschild-Stra{\ss}e 1, 85741 Garching, Germany}\\
\normalsize{$^{10}$Subaru Telescope, National Astronomical Observatory of Japan (NAOJ),}\\
\normalsize{650 North A'ohoku Place, Hilo, HI 96720}\\
\normalsize{$^{11}$National Astronomical Observatory, CAS, 20A Datun Road, 
Chaoyang District, }\\
\normalsize{Beijing 100012, China}\\
\normalsize{$^{12}$Division of Optical and Infrared Astronomy, NAOJ, Osawa 2-21-1, Mitaka, }\\ 
\normalsize{Tokyo 181-8588, Japan}\\
\normalsize{$^{13}$National Optical Astronomy Observatory, Tucson, AZ 85719, USA}\\
\normalsize{$^{14}$Department of Astronomy, University of California, Berkeley, 
CA 94720-3411, USA}\\
\normalsize{$^{15}$ALMA Project, NAOJ, Mitaka, Tokyo 181-8588, Japan}\\
\normalsize{$^{16}$Department of Infrared Astrophysics, ISAS,
   Japan Aerospace Exploration Agency (JAXA), }\\
\normalsize{3-1-1 Yoshinodai, Sagamihara, Kanagawa 229-8510, Japan}\\
\normalsize{$^{17}$Astronomy Data Center, NAOJ, Mitaka, Tokyo 181-8588, Japan}
\\
\normalsize{$^\ast$To whom correspondence should be addressed; E-mail:
maeda@ea.c.u-tokyo.ac.jp} 
}

\date{}

\begin{document} 




\maketitle


\begin{sciabstract}
Core-collapse supernovae (CC-SNe) are the explosions that announce the death of 
massive stars. 
Some CC-SNe are linked to long-duration gamma-ray bursts (GRBs) and are highly
aspherical. 
One important question is to what extent asphericity is common to all CC-SNe. 
Here we present late-time spectra for a number of CC-SNe from stripped-envelope
stars, and use them to explore any asphericity generated in the inner part of 
the exploding star, near the site of collapse. 
A range of oxygen emission-line profiles is observed, including a high 
incidence of double-peaked profiles, a distinct signature of an aspherical 
explosion. 
Our results suggest that all CC-SNe from stripped-envelope stars are aspherical
explosions and that SNe accompanied by GRBs exhibit the highest degree of 
asphericity.
\end{sciabstract}

Massive stars ($M \gsim 10 \Msun$) end their lives when the nuclear fuel in 
their innermost region is consumed; lacking sufficient internal pressure
support, they can no longer withstand the pull of gravity.  Their core then
collapses to a neutron star or a black hole.  Gravitational energy from the
collapse produces an explosion that expels the rest of the star in what is
observed as a supernova (SN). 

Core-collapse SNe (CC-SNe) are classified ({\it 1}) by how much of the stellar
envelope is present at the time of the explosion ({\it 2}).  Stars that retained
their hydrogen envelope produce SNe with a H-rich spectrum, classified as Type
II.  On the other hand, stars that have lost all, or most, of the H-envelope
produce CC-SNe that are known as ``stripped-envelope" (briefly ``stripped'') SNe.
These include, in a sequence of increasing envelope stripping, Type IIb (He
rich, but still showing some H), Type Ib (He-rich, no H), and  Type Ic (deprived
of both H and He).  Some SNe\,Ic (hereafter ``broad-lined" SNe\,Ic) show very
broad absorption features in optical spectra obtained within a few weeks of the
explosion; these features are produced by material moving at $v > 0.1c$ ($c$ is
the speed of light), probably the result of an explosion characterized by a
kinetic energy ($E_{\rm K}$) larger than the canonical value of $\sim 10^{51}$
erg ({\it 3}). The most energetic broad-lined SNe\,Ic can reach $E_{\rm K}
\gsim 10^{52}$ erg (hereafter ``hypernovae" or ``GRB-HNe") ({\it 4}), and can be
associated with gamma-ray bursts (GRBs) ({\it 5}).  Figure 1 summarizes the
relation between $E_{\rm K}$ and the mass of $^{56}$Ni that powers the optical
light of stripped CC-SNe ({\it 6}). 

An important unsolved question concerns how gravitational energy of the collapse
is turned into outward motion of the SN explosion.  Most recently proposed
scenarios involve aspherical explosions ({\it 7--11}).  Therefore, mapping the
explosion geometry can be illuminating.  It is especially critical to establish 
whether the explosion geometry is similar for all CC-SNe, or at least for
different subclasses (GRB-HNe, broad-lined SNe\,Ic, normal-energy SNe, or
stripped vs. Type II SNe). 

Some evidence for asymmetric explosions was obtained from the polarization
detected in several SNe~II ({\it 12}) and SNe~Ib/c ({\it 13}), and in
the broad-lined SN~Ic 2002ap ({\it 14--16}).  However,
as there are still few such detections, no systematic study exists. 

The best way to investigate a supernova's inner ejecta geometry is through 
late-time spectroscopy.  At $t \gsim 200$ days after the explosion, expansion
makes the density of the ejecta 
so low that optical photons produced anywhere in the ejecta
escape without interacting with the gas.  At these epochs, the SN spectrum is
nebular, showing emission lines mostly of forbidden transitions. Because the
expansion velocity is proportional to radius of any point in the ejecta, 
the Doppler shift indicates where the
photon was emitted: a photon emitted from the near/far side of the ejecta is
detected at a shorter/longer (blueshifted/redshifted) wavelength.  The late-time
nebular emission profiles thus probe the distribution of the emitting gas within
the SN ejecta.  This strategy is particularly effective for stripped CC-SNe,
since we can look directly into the oxygen core. 

Analysis of the late-time spectra of the GRB-HN SN\,Ic\,1998bw ({\it 17})
and of the broad-lined SN~Ic 2003jd ({\it 18}) provided evidence that
these objects shared a similar, bipolar explosion.  However, we viewed
SN\,1998bw on-axis, and SN\,2003jd sideways.  This seems consistent with the
fact that SN\,1998bw was associated with a GRB, while SN\,2003jd was not, and 
suggests that if SN\,2003jd also produced a GRB, this was missed because of
its orientation (({\it 18}), but see ({\it 19}) for concerns). 

We obtained late-time spectra of stripped CC-SNe to study their morphology and
quantify their properties.  Our data were obtained mostly with FOCAS ({\it 20})
on the 8.2\,m Subaru telescope and with FORS2 on the ESO Very Large Telescope
(VLT).  Additional data  are from ({\it 21}).  The strongest emission line in
stripped CC-SNe is [\OI] $\lambda\lambda$6300, 6363. Despite being a doublet, it
behaves like a single transition if the lines are sufficiently broad ($\gsim 0.01 c$),
because the red component is weaker than the blue one by a factor of 3 
(see supporting online text). 

Since at epochs larger than 200 days after the explosion the ejecta are 
transparent to line emission, and radiation transfer is unimportant (see supporting
online text), we selected spectra obtained at least 200 days after discovery to
build an unbiased data set.  Additionally, we did not include hypernovae
discovered through an associated GRB, to avoid bias in the viewing angle.  The
selection procedure and possible biases are discussed in ({\it 22}).  Our sample
(Tab. S1) is the largest published data set to date of stripped CC-SNe at such
late epochs.  Figure 2 shows the spectra of the 18 SNe in our sample. 
Among them, 13 are presented here for the first time. 

The observed [\OI] $\lambda\lambda$6300, 6363 emission profiles can be
compared with the prediction of various explosion models.  We use three
representative models from ({\it 23}): one extremely aspherical (BP8), 
one mildly aspherical (BP2), and one spherical (BP1). 

In the spherical model, $^{56}$Ni is confined in a central high-density region
with an inner hole, and is surrounded by a low-density, oxygen-rich region 
({\it 24}).  This results in a single-peaked, but flat-topped [\OI]
profile, independent of the orientation. 

On the other hand, the bipolar model ({\it 24--26}) is characterized by a
low-density $^{56}$Ni-rich region located near the jet axis, where the jets
convert stellar material (mostly oxygen) into Fe-peak elements, and by a
high-density disk-like structure composed of unburned oxygen-rich material, as
the jet expands laterally only weakly ({\it 17, 24}) (see Fig. S1).  The [\OI]
profile in a bipolar model depends on both the degree of asphericity and the
viewing angle ({\it 23, 27}) (see supporting online text and Fig. S1).  If a
bipolar supernova explosion is viewed from a direction close to the jet axis,
the O-rich material in the equatorial region expands in a direction
perpendicular to the line of sight, and the [\OI] emission profile is observed
to be sharp and single-peaked. On the other hand, for a near-equatorial view the
profile is broader and double-peaked.  The best fit to the light curve and the
spectra of the GRB-HN SN 1998bw was obtained with model BP8 ({\it 23, 27}). 

For the degree of asphericity of this model, the [\OI] profile should switch
from single- to double-peaked near a viewing angle $\theta \approx 50^{\circ}$
measured from the jet direction (Fig. S2). The predicted frequency of
double-peaked [\OI] is thus $\sim 64$\%, in the absence of bias in the
orientation.  For a less aspherical model the fraction of double peaks is
reduced: model BP2 shows double peaks only for $\theta \gsim 70^{\circ}$,  and
has a double-peak fraction of $\sim 34$\%. With this variety of [\OI] profiles, 
statistics (Tab. 1) can be used to constrain the degree of asphericity and
remove the uncertainty in the viewing angle. 

Figure 3 shows the [\OI] emission profiles for our sample.  Out of 18 SNe, 5
(SNe\,2003jd, 2004fe, 2005aj, 2005kl, and 2006T) clearly show double-peaked
[\OI] profiles.  Four others (SNe\,1997dq, 2004gn, 2005kz, and 2005nb) are
apparently transitional objects with flat-topped or mildly peaked [\OI]
profiles, and there is 
a marginal detection of double peaks in some cases.  The remaining
9 SNe exhibit single-peaked profiles.  For illustrative purposes, the observed
profiles are compared with model predictions in Figure 3. 

The line profiles are well reproduced by the bipolar explosion model assuming
different viewing angles.  Although detailed fits are not unique for individual
objects, this uncertainty does not affect the bulk statistics (single- vs.
double-peaked): the presence of double-peaked [\OI] profiles is not  predicted
in spherical models, and their fraction yields a secure estimate of the number
of aspherical SNe viewed sideways, assuming that the sample is unbiased in
orientation.  The high incidence of double-peaked profiles is an important
discovery: double-peaked [\OI] was previously reported only for SN\,Ic 2003jd 
({\it 18}) and SN\,Ib 2004ao ({\it 28}).  The observed fraction of double-peaked
profiles, $(5~{\rm to}~9)/18$ (= 28\%--50\%, median $\sim 39$\%) is consistent
with all stripped CC-SNe being mildly aspherical, like model BP2. 

On the other hand, the observed fraction does not support the possibility that 
all stripped CC-SNe are extremely aspherical explosions like SN 1998bw (i.e.,
model BP8), as then we would expect an even larger fraction of double-peaked
[\OI] profiles: according to our Monte Carlo simulations with randomly oriented
viewing directions, the observed number should be $7-16$ out of 18 SNe with 99\%
confidence level and $10-13$ with 70\% confidence level.  Alternatively, about
half of all stripped CC-SNe may have asphericity as large as that of GRB-HNe,
with a double-peak incidence of $\sim 64$\% (BP8), with the remaining half being
approximately spherical and yielding the bulk of the single-peaked profiles. 
With a larger SN sample we could look in more detail at [\OI] profiles as
functions of the degree of asphericity and orientation, and more fully explore 
these scenarios. 

Although our sample is still small, we can look for statistical differences
between GRB-HNe, broad-lined SNe\,Ic, and other stripped CC-SNe. In our sample,
six SNe are broad-lined (non-GRB) SNe\,Ic (SNe\,1997ef, 1997dq, 2002ap, 2003jd,
2005kz, and 2005nb), while the rest are probably ``normal" stripped CC-SNe.  
The observed fraction of double-peaked [\OI] is 33\% (1 out of 3 SNe) for
broad-lined SNe\,Ic and 36\% (4 out of 11) for the others, if SNe with
transitional [\OI] profiles (Fig. 3) are excluded. 
Within the statistical uncertainty, 
caused especially by the small sample of broad-lined SNe\,Ic, there is no
difference between the two groups.  Both are consistent with the predictions of
model BP2, and show too few double-peaked profiles for model BP8.  This suggests 
that on average, broad-lined, non-GRB SNe\,Ic are less aspherical than GRB-HNe, 
and morphologically closer to normal stripped CC-SNe. 

Broad-lined, non-GRB SNe\,Ic have typically smaller $E_{\rm K}$ than GRB-HNe
(Fig. 1).  There has been speculation that broad-lined, non-GRB SNe\,Ic might be
intrinsically similar to GRB-HNe but viewed off-axis, leading to apparently
smaller $E_{\rm K}$.  Although this may still be true for a small subset of
them, the moderate asphericity inferred for this group suggests that
broad-lined, non-GRB SNe\,Ic are probably intrinsically different from GRB-HNe.

On the other hand, asphericity is not a special feature of GRB-HNe, but rather a
generic property of stripped CC-SNe.  Both broad-lined and normal stripped
CC-SNe have a moderate degree of asphericity.  All stripped CC-SNe probably
share to some extent a common explosion mechanism that generates the same kind
of asphericity, with GRB-HNe likely the most aspherical.  
Our result offers an 
important insight into the theory of SN explosions.  The most popular models for
high- and low-energy CC-SNe are in fact different: black hole formation and the
production of a jet in hypernovae and perhaps broad-lined SNe\,Ic ({\it 29});
delayed neutrino heating from the proto-neutron star for normal SNe ({\it 30}). 
In the former case, the explosion is probably initiated along the axis of
rotation and/or magnetic field ({\it 10, 11, 29}), while in the latter case some
asphericity may be generated by hydrodynamic instabilities ({\it 7--9}).  
The result supports recent theoretical scenarios of the supernova explosion, which 
suggest that an important role in the collapse is played by hydrodynamic 
instability, rotation, or magnetic fields.

Recently, Modjaz et al. ({\it 31}) showed 
another sample of late-time spectra of stripped CC-SNe. 
Their conclusions are similar to ours. 
One SN in their sample (SN 2004ao with double-peaked [\OI]) 
could be added to our sample according to our selection criteria.
It increases the frequency of double-peaked events, but it
does not change our conclusions within the uncertainties. 

\subsection*{References and Notes}

\begin{itemize}

\item[1.]
A. V. Filippenko,
{\it Ann. Rev. Astron. Astrophys.} {\bf 35}, 309 (1997). 

\item[2.]
K. Nomoto, K. Iwamoto, \& T. Suzuki,
{\it Phys. Rep.} {\bf 256}, 173 (1995). 

\item[3.]
P. A. Mazzali, et al.,
{\it Astrophys. J.} {\bf 572}, L61 (2002). 

\item[4.]
K. Iwamoto, et al.,
{\it Nature} {\bf 395}, 672 (1998).

\item[5.]
T. J. Galama, et al., 
{\it Nature} {\bf 395}, 670 (1998). 

\item[6.]
K. Nomoto, et al., 
{\it Il Nouvo Cimento} {\bf 121B}, 1207 (2007) 
(astro-ph/0702472).

\item[7.]
J. M. Blondin, A. Mezzacappa, \& C. DeMarino,
{\it Astrophys. J.} {\bf 584}, 971 (2003). 

\item[8.]
R. Buras, H.-Th. Janka, M. Rampp, \& K. Kifonidis, 
{\it Astronomy and Astrophysics} {\bf 457}, 281 (2006). 

\item[9.]
A. Burrows, E. Livne, 
L. Dessart, C. D. Ott, \& J. Murphy 
{\it Astrophys. J.} {\bf 655}, 416 (2007). 

\item[10.]
K. Kotake, H. Sawai, 
S. Yamada, \& K. Sato,
{\it Astrophys. J.} {\bf 608}, 391 (2004). 

\item[11.]
S. G. Moiseenko, G. S. Bisnovatyi-Kogan, \& 
N. V. Ardeljan,
{\it Mon. Not. Roy. Astron. Soc.} {\bf 370}, 501 (2006). 

\item[12.]
D. C. Leonard, et al., 
{\it Nature} {\bf 440}, 505 (2006). 

\item[13.]
L. Wang, D. A. Howell, P. H\"oflich, \& J. C. Wheeler,
{\it Astrophys. J.} {\bf 550}, 1030 (2001). 

\item[14.]
K. S. Kawabata, et al.,
{\it Astrophys. J.} {\bf 580}, L39 (2002). 

\item[15.]
D. C. Leonard,  A. V. Filippenko, R. Chornock, \& R. J. Foley, 
{\it Publ. Astron. Soc. Pacific} {\bf 114}, 1333 (2002).
 
\item[16.]
L. Wang, D. Baade, P. H\"oflich, \& J. C. Wheeler,
{\it Astrophys. J.} {\bf 592}, 457 (2003). 

\item[17.]
K. Maeda, et al., 
{\it Astrophys. J.} {\bf 565}, 405 (2002). 

\item[18.]
P. A. Mazzali, et al.,
{\it Science} {\bf 308}, 1284 (2005). 

\item[19.]
A. M. Soderberg, E. Nakar, E. Berger, \& S. R. Kulkarni,
{\it Astrophys. J.} {\bf 638}, 930 (2006). 

\item[20.]
N. Kashikawa, et al., 
{\it Publ. Astron. Soc. Japan} {\bf 54}, 819 (2002). 

\item[21.]
T. Matheson, A. V. Filippenko, 
W. Li, D. C. Leonard, \& J. C. Shields, 
{\it Astronomical J.} {\bf 121}, 1648 (2001). 

\item[22.]
Data and Methods are available as supporting online materials 
on Science Online. 

\item[23.]
K. Maeda, K. Nomoto, P. A. Mazzali, \& J. Deng, 
{\it Astrophys. J.} {\bf 640}, 854 (2006). 

\item[24.]
K. Maeda \& K. Nomoto, 
{\it Astrophys. J.} {\bf 598}, 1163 (2003). 

\item[25.]
A. M. Khokhlov, P. A. H\"oflich, E. S. Oran, 
J. C. Wheeler, L. Wang, \& A. Yu. Chtchelkanova, 
{\it Astrophys. J.} {\bf 524}, L107 (1999). 

\item[26.]
A. I. MacFadyen, S. E. Woosley, \& A. Heger, 
{\it Astrophys. J.} {\bf 550}, 410 (2001). 

\item[27.]
K. Maeda, P. A. Mazzali, \& K. Nomoto, 
{\it Astrophys. J.} {\bf 645}, 1331 (2006). 

\item[28.]
M. Modjaz, et al.,
{\it Astron. J.} {\bf in press} (2008) 
(astro-ph/0701246). 

\item[29.]
A. I. MacFadyen \& S. E. Woosley, 
{\it Astrophys. J.} {\bf 524}, 262 (1999). 

\item[30.]
H. A. Bethe \& J. R. Wilson, 
{\it Astrophys. J.} {\bf 295}, 14 (1985). 

\item[31.]
M. Modjaz, R.P. Kirshner, P. Challis, preprint (2008) 
(astro-ph/0801.0221)

\item[32.]
The data were collected at the Subaru Telescope 
(the National Astronomical Observatory of Japan). 
The additional data were from 
the Very Large Telescope (European Southern Observatory) 
under ESO Program 078.D-0246. 
K.M. and M.T. have been supported by the Japan Society for the Promotion 
of Science (JSPS). 
E.P. and P.M. acknowledge financial support from PRIN MIUR 2005 
and from PRIN INAF 2006. 
K.N. is supported by the JSPS Grant-in-Aid for Scientific Research (18104003, 18540231).
J.D. is supported by NSFC (No. 10673014). 
A.V.F. is supported by NSF grant AST--0607485. 

\end{itemize}

\clearpage

\clearpage
\begin{figure}
	\begin{minipage}[t]{0.48\textwidth}
		\psfig{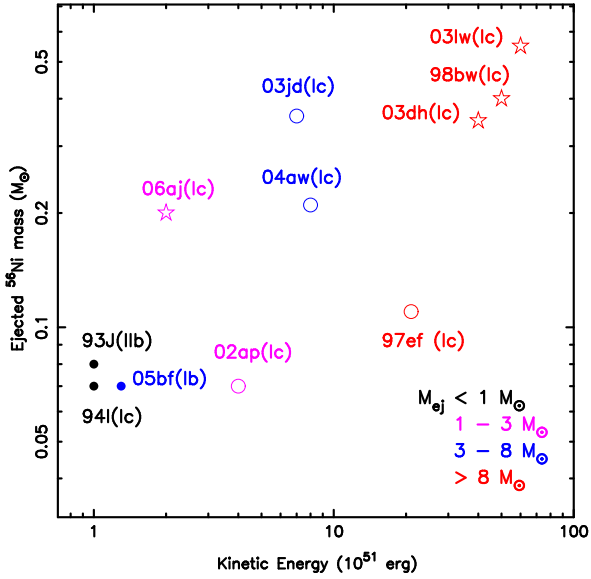}
	\end{minipage}
\caption{Relation between the kinetic energy of the explosion ($E_{\rm K}$) 
and the mass of ejected $^{56}$Ni [$M$($^{56}$Ni)] of stripped CC-SNe 
(see supporting online text). 
Colors indicate the ejecta mass ($M_{\rm ej}$). 
For SN IIb 1993J and SN Ib 2005bf, the ejecta mass after subtracting the He 
envelope mass is shown as $M_{\rm ej}$, to compare with SNe Ic, 
which lack the He envelope. 
SNe associated with GRBs or an X-ray flash (XRF) are indicated by stars, 
and broad-lined SNe Ic without GRBs/XRFs by open circles. 
Other (normal) stripped CC-SNe are shown as dots. 
}
\end{figure}

\clearpage
\begin{figure}
\psfig{file=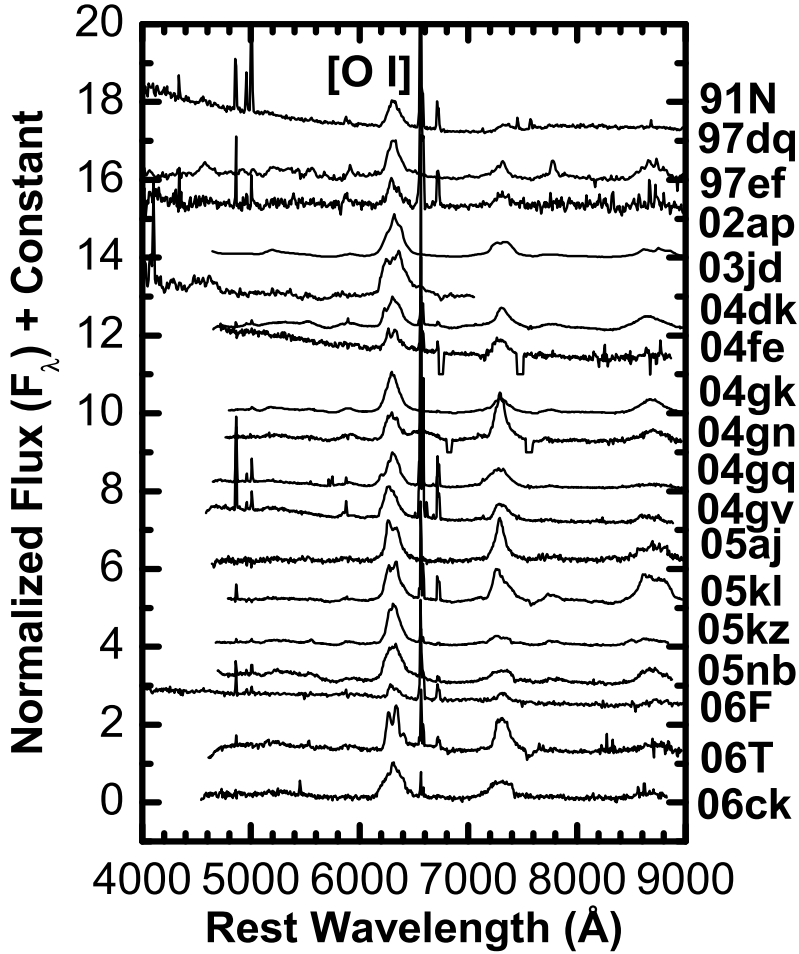,width=0.8\textwidth}
\caption{Nebular spectra of the supernovae in the sample used in this study. 
Narrow lines (e.g., H${\alpha}$ at 6563\,\AA) originate from a diffuse 
superposed H~II region, not from a SN. 
The spectra of SNe 1991N, 1997dq, 1997ef are from ({\it 21}). 
The other spectra were obtained with the Subaru telescope, except for 
SN 2006F which was taken by the VLT ({\it 22}). 
Spectra were deredshifted using the redshift obtained from the
observed wavelength of the narrow H${\alpha}$ emission if possible; 
otherwise, the redshift of the nucleus of the host galaxy was adopted.
For presentation, the flux is arbitrarily scaled and shifted vertically. 
The strongest emission line (dashed line) is [\OI] $\lambda\lambda$6300, 6363. 
The feature at $\sim 7,300$\,\AA\ is [\CaII] $\lambda\lambda$7291, 7324 
contaminated by several emission lines: [\OII], [\FeII], and [\NiII].  
}
\end{figure}

\clearpage
\begin{figure}
	\begin{minipage}[t]{0.48\textwidth}
		\psfig{file=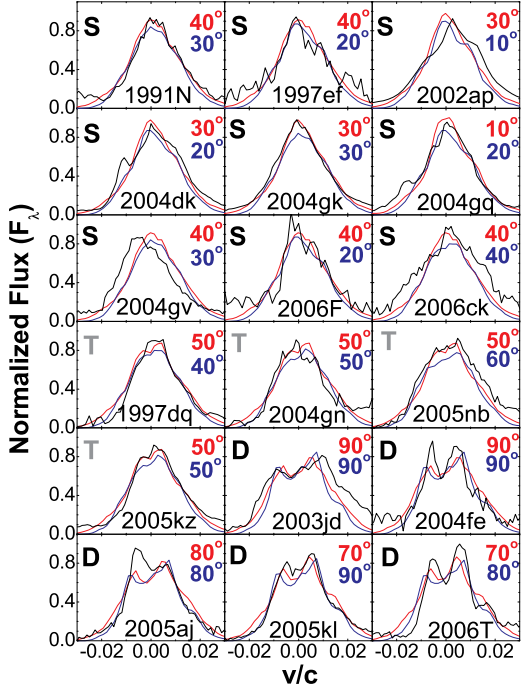,width=1.0\textwidth}
	\end{minipage}
\caption{Observed [\OI] $\lambda\lambda$6300, 6363 emission-line profiles 
(black curves), classified into characteristic profiles: single peaked (denoted
by ``S," top 9 SNe), transition (``T," middle 4 SNe), and double peaked (``D,"
bottom 5 SNe).  For presentation, model predictions of the bipolar
model ({\it 23}) with  different viewing directions are shown for Model
BP8 (red curves, with the direction denoted by the red text) and for the less
aspherical model BP2 (blue).  The models shown here have smaller expansion
velocities (corresponding to $E_{\rm K} \approx $ a few $\times 10^{51}$ erg) 
than the one applied to SN\,1998bw ($E_{\rm K} \gsim 10^{52}$ erg) ({\it 23}).  The
separation between the two peaks in the double-peaked cases is $\sim 0.01 c -
0.02 c$, much larger than the wavelength resolution in the observations ($\sim
10$\,\AA, corresponding to $0.0015 c$). 
}
\end{figure}

\clearpage
\begin{table}
\caption{Fraction of double-peaked [\OI] SNe.}
\begin{center}
\begin{tabular}{ccc}
\hline
Model & Dividing angle\footnotemark[1] & Fraction\footnotemark[2]\\
\hline\hline
Spherical & --- & $0$\%\\
BP2       & $\sim 70^{\circ}$ & $\sim 34$\%\\
BP8       & $\sim 50^{\circ}$ & $\sim 64$\%\\
\hline
Observed  & ---          & $39 \pm 11$\%\\
\hline
\end{tabular}
\end{center}
\footnotetext{}{$^{1}$The viewing angle ($\theta_{0}$ measured from the jet 
direction) which divides the [\OI] profile --- i.e., single peaked if 
$\theta < \theta_{0}$ and double peaked if 
$\theta_{0} \le \theta \le 90^{\circ}$.\\
$^{2}$Fraction of SNe showing double-peaked [\OI].}
\end{table}

\clearpage

\begin{center}
\vspace*{-10.00pt}
{\Large \bf SUPPORTING ONLINE MATERIAL}
\end{center}
\begin{center}
{\large Asphericity in Supernova Explosions 
from Late-Time Spectroscopy}\\
\end{center}
\begin{center}
{\large Keiichi Maeda et al.}
\end{center}

\subsection*{Data and Methods}

\noindent{\bf Sample of Supernova Nebular Spectra: }\\
In the years 2005 to early 2007, we 
obtained a number of spectra of stripped CC-SNe 
using the 8.2~m Subaru telescope 
equipped with FOCAS. We also 
obtained several spectra using the VLT with FORS2 
from late 2006 to early 2007. 
Some of the spectra are distinctly nebular, while others are not. 
In addition to this new sample, 
we include some spectra presented in ({\it S1}), 
which contains the largest sample of previously published 
late-time spectra of stripped CC-SNe. 
The spectral resolution is $\sim 10$~\AA\ for the spectra 
obtained by the Subaru or VLT. 
The spectra from the previous sample ({\it S1}) 
were obtained with the Shane 3-m reflector at Lick Observatory, with 
similar spectral resolution (6--15~\AA). 

Because of the faintness of SNe in the late phases, 
the spectra can be obtained only for relatively bright 
SNe even with the largest telescopes in the world. 
The limiting functions are thus the intrinsic luminosity 
and the distance. The luminosity bias may be important 
when interpreting the data if the effect is strong: 
our sample then represents a subclass of luminous SNe (i.e., 
SNe with a large amount of $^{56}$Ni), and may affect the 
asymmetry statistics we investigate in the present paper. 

However, we believe that the effect is not strong enough to affect 
our interpretation. First, our criterion for successful 
spectroscopy is $\sim 23$ mag. This roughly corresponds 
to SNe with a peak magnitude of 17 or 18. This 
brightness is similar to the criterion for obtaining high-quality 
data with typical SN observations in the early phases 
using smaller telescopes. As such, our sample should not 
be very different from samples of well-studied stripped CC-SNe 
in past studies using early-phase observations which, 
for example, led to the result shown in Figure 1. 
Second, the peak luminosity of very luminous stripped CC-SNe is 
larger than that of typical ones by at most a factor of $\sim 5$ 
(e.g., SN 1998bw ({\it S2})), so
the detection volume of these luminous events is 
larger than that of typical ones by at most one order of magnitude. 
However, the intrinsic rate of the very luminous events is 
probably smaller than that of the typical ones by nearly two 
orders of magnitude ({\it S3}). Thus, contamination 
by very luminous SNe should not be large in our sample. 
Finally, there is a correlation between the $^{56}$Ni mass 
and the early-phase width of absorption lines ({\it S4}). Using this 
relation, we indeed divide our sample into a potentially luminous 
class (i.e., broad-lined SNe Ic) and 
the other ones in the discussion of the asphericity statistics. 
Also, according to past studies, normal stripped CC-SNe (i.e., not broad-lined SNe~Ic) 
have similar $^{56}$Ni mass ($\sim 0.1\Msun$) and represent a wide range
of progenitor masses up to $\sim 20\Msun$ ({\it S4}). This suggests that 
the luminosity bias, if present, is not strong in the normal stripped 
CC-SNe. In summary, the luminosity bias should not be strong in 
our sample of late-time spectra. 
Thus, the distance is probably the limiting factor,
and our data consist of SNe within a distance 
of at most $\sim 100$ Mpc.

Other possible biases are (1) a possibility of 
preferentially observing SNe from a special direction 
(e.g., a jet axis), and (2) contamination of SNe which are young and 
dense so that their ejecta are not sufficiently transparent 
to trace the geometry by the [\OI] profile. 
Here we describe how we selected 
the final unbiased list of supernova spectra 
expected to be at a fully nebular epoch to study 
the intrinsic line profile (Table S1). 

First, we excluded SNe 1998bw, 2003dh, 2005bf, and 2006aj.
SN~Ic 2006aj was associated with an X-Ray flash 
(XRF: a low-energy analog of a GRB) ({\it S5});
thus, the detection of SN 2006aj is probably biased by the viewing-angle 
effect, being easier for an observer along the jet direction. 
For the same reason, we excluded SNe 1998bw and 2003dh, associated with GRB 
980425 and GRB 030329, respectively, although there are some 
published late-phase observations for these objects ({\it S6, S7}). 
SN Ib 2005bf is reported to be a very peculiar SN, whose 
explosion mechanism may be different from that of typical CC-SNe;
the nebular observations are found to be peculiar, too ({\it S8}). 
Observations and analyses of SN 2006aj and SN 2005bf 
in nebular phases are described 
in separate papers ({\it S8--10}). 
Note that in the late-time phases considered here, the dependence of 
luminosity on the viewing angle is negligible. 

Next, we excluded supernovae which were observed only at 
times before entering the fully nebular phase, since
optical transport effects can change the line profile. 
The choice is not trivial; the optical depth ($\tau$) 
differs for different supernovae ($\tau \propto 
M_{\rm ej}^2/E_{\rm K}$), and the ejecta mass ($M_{\rm ej}$) 
and the ejecta kinetic energy ($E_{\rm K}$) 
are not known a priori. 
The uncertainty in the explosion date further complicates matters.

Therefore, we applied a very simple criterion.  We checked the 
phases of the observations with respect to the discovery date ($t$), 
and selected a supernova only if it was observed 
at $t > 200$ days at least once. 
Thus, any SNe possibly younger than $200$ days 
were excluded from the list. 
This criterion, $ t > 200$ days, is 
sufficiently late so that any selected supernova is certainly in the 
nebular phase. 

Following these procedures, we obtained the unbiased sample 
of stripped CC-SNe whose spectra are distinctly nebular. 
The sample contains 18 SNe. One of the SNe is of Type IIb. 
The sample contains 13 SNe newly observed 
(11 only by the Subaru telescope, and 2 both by the Subaru and 
VLT). Spectra of two broad-lined SNe Ic (2002ap and 2003jd) 
were also obtained with the Subaru telescope, which 
were already reported ({\it S11, S12}). 
The remaining 3 SNe are from the previously published sample of 
({\it S1}). 
In terms of the spectral characteristics in the early phase, 
6 SNe in our list have been classified as broad-lined SNe Ib/c 
without association with a GRB (SNe 1997dq, 1997ef, 
2002ap, 2003jd, 2005kz, and 2005nb -- see also Figure 1 in the main text). 

In cases where the supernova was observed at multiple epochs, we included the 
object in the nebular sample if the last observation 
satisfied the condition $t > 200$ days. Such SNe were used to check 
the validity of the selection criterion. 
SN 2005aj was observed at three epochs ($t = 189$, $249$, and 
$312$ days), showing significant evolution of the [\OI] $\lambda\lambda$6300, 
6363 profile between the first two epochs. 
SN 2006F was observed at $t = 173$ and $312$ days, showing 
no noticeable evolution in the [\OI] profile. 
The other SNe observed multiple times at $t \gsim 200$ days 
do not show noticeable evolution in the [\OI] profile. 
These observations justify our criterion ($t > 200$ days). 

All the spectra shown in Figures 2 and 3 in the main text 
are at $t > 200$ days except for that of SN 2005nb. 
The signal-to-noise ratio for the spectrum of SN 2005nb at $t = 403$ days 
is low, but the [\OI] profile does not evolve significantly from 
$t = 195$ days (shown in Figures 2 and 3 in the main text).

\subsection*{Supporting Text}

\noindent{\bf The properties of stripped CC-SNe: $E_{\rm K}$, $M$($^{56}$Ni), and $M_{\rm ej}$: }\\
The kinetic energy of the explosion ($E_{\rm K}$), the mass of $^{56}$Ni [$M$($^{56}$Ni)], 
and the ejecta mass ($M_{\rm ej}$) shown in Figure 1 of the main text are 
derived by modeling the early-phase observations using approximate analytical 
expressions (SNe 2003jd and 2004aw) or one-dimensional radiation transport 
codes (the other SNe). 
An exception is SN Ib 2005bf, for which $M$($^{56}$Ni) is derived by 
modeling the late-time observations ({\it S8}). 

Assuming spherical symmetry, the estimate of $E_{\rm K}$ (shown in Figure 1) 
represents an isotropic value, which could be different from the intrinsic value
by up to a factor of a few.  For example, for SN 1998bw (associated with GRB
980425), we estimated $E_{\rm K} \approx 5 \times 10^{52}$ erg with a spherical
model ({\it S2}), but $E_{\rm K} \approx 2 \times 10^{52}$ erg with a
two-dimensional jet model ({\it S13}).  The difference can be attributed
to a viewing-angle effect.  Since SN 1998bw is an extreme case (see the main
text), the difference between the isotropic value and the real one should be
smaller for other SNe. 

The references are as follows: 
SN IIb 1993J ({\it S14}), SN Ic 1994I ({\it S15}), 
SN Ic 1997ef ({\it S16}), 
SN Ic 1998bw ({\it S2}),
SN Ic 2002ap ({\it S17}), SN Ic 2003dh ({\it S18}), 
SN Ic 2003jd ({\it S19}), SN Ic 2003lw ({\it S20}), 
SN Ic 2004aw ({\it S19}), SN Ib 2005bf ({\it S8, S21}), 
and SN Ic 2006aj ({\it S22}). \\

\noindent{\bf The Aspherical SN Model: }\\
The aspherical models presented in this paper 
are from ({\it S23, S24}). 
The jet-driven explosion is simulated by promptly 
depositing the kinetic and thermal energy in the innermost core. 
The initial kinetic energy is distributed in a jet-like way, applying 
$v_{z} = \alpha z$ and $v_{r} = \beta r$, where 
($r$, $z$) is a cylindrical coordinate system (assuming axial symmetry 
with respect to the $z$-axis), and $v_{z}$ and $v_{r}$ 
are the initial deposited velocity components. 
Models BP8 and BP2 assume $\alpha/\beta = 8$ and $2$, respectively; 
see ({\it S23, S24}) for details. 
The distributions of density, $^{56}$Ni, and $^{16}$O are shown in Figure S1. 

The optical depth to the [\OI] $\lambda$6300 can 
be approximated by $0.01 M_{\rm O}/\Msun (t_{\rm exp}/200 \ \ {\rm days})^{-2}$ 
for a homogeneous medium with an oxygen mass of $M_{\rm O}$ 
and an expansion velocity of $0.02 c$ (where $t_{\rm exp}$ is 
time since the explosion) ({\it S25}). The [\OI] $\lambda$6363 
line is always more transparent than the 6300~\AA\ component 
because the latter has a larger transition probability.
Thus, the effect of the optical depth to the [\OI] line profile should be 
negligible. 
Furthermore, we have confirmed that the [\OI] line 
is always transparent in our models at $t_{\rm exp} > 200$ days. 
In this optically thin case, the blue (6300~\AA) component is stronger than 
the red (6363~\AA) one by a factor of 3, and the doublet 
behaves like a single transition. 

The jet-driven model predicts unique features in the [\OI] profile 
(Figs. S1 and S2) ({\it S23, S24}). 
The SN ejecta kinematics are described by homologous expansion, 
$R = v (R) t_{\rm exp}$, where $R$ and $v(R)$ are respectively the radius 
(in a spherical coordinate) and the radial expansion velocity of a 
point in the ejecta.
Non-radial velocity components should be negligible.  
The observed wavelength of a photon depends on the point from which it was
emitted, as $\lambda = \lambda_0 (1 - v_{||}/c)$, where $\lambda_0$ is the rest 
wavelength of the line, $v_{||}$ is the line-of-sight velocity toward the 
observer, and $c$ is the speed of light. 
Then, defining $d$ as the projection of $R$ onto the line of sight, 
it can be shown that $v_{||} = d / t_{\rm exp}$, and this is the same 
in a given plane perpendicular to the line of sight. 
This means that, in the late phase, all photons emitted 
at the same depth along the line-of-sight have the same wavelength. 
The wavelength of a photon emitted from the near/far side of
the ejecta is detected as shorter/longer (blueshifted/redshifted).
Thanks to these characteristics, and to the transparency of the ejecta, 
the late-time emission profile can be used to probe the density distribution 
of the emitting gas within the SN ejecta (the ``SN scan''), just like 
Computerized Tomography scans the human body (the ``CT scan''). 
The SN scan is shown in Figure S1 for observers along the 
$z$-direction and $r$-direction, respectively. 
The model prediction (BP8) for the [\OI] profile is shown in 
Figures S1 and S2 as a function of the viewing angle ($\theta$). 

If we view the bipolar supernova explosion from a direction close to the jet 
axis, most of the O-rich material is in the equatorial region, expanding in a 
direction perpendicular to the line of sight and producing a narrow-peaked [\OI] 
emission profile (Fig. S2). 
The SN scan provides a monotonically increasing amount of O-rich material
as one probes nearer the center of the SN, then a monotonically decreasing 
amount of O-rich material as one moves away from the center (Fig. S1). 
On the other hand, the [\OI] line shows a double-peaked profile 
if viewed from near the equatorial direction. 
In this case, the SN scan results in an increasing amount of O-rich material as
one moves from the observer to the edge of the inner hole of the O-rich disk, 
then a decrease toward the center of the SN, where the non-emitting hole 
occupies the largest area. The amount of O then again increases as one moves 
from the center to the edge of the hole on the other side, and subsequently it
again decreases toward the receding side of the ejecta (Fig. S1).  
Thus, two characteristic Doppler-shifted emission peaks 
(blueshifted and redshifted) appear.

\clearpage
\begin{figure*}
\begin{minipage}[t]{0.45\textwidth}
		\psfig{file=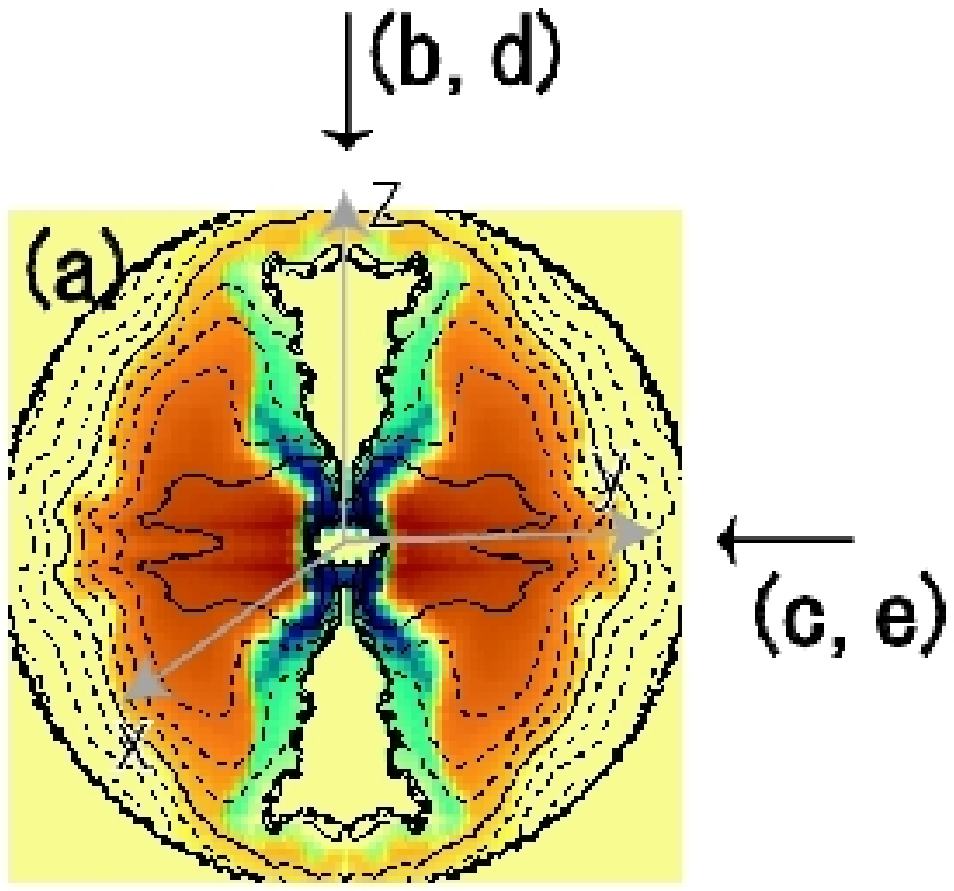,width=1.0\textwidth}
	\end{minipage}\\
	\begin{minipage}[t]{0.45\textwidth}
		\psfig{file=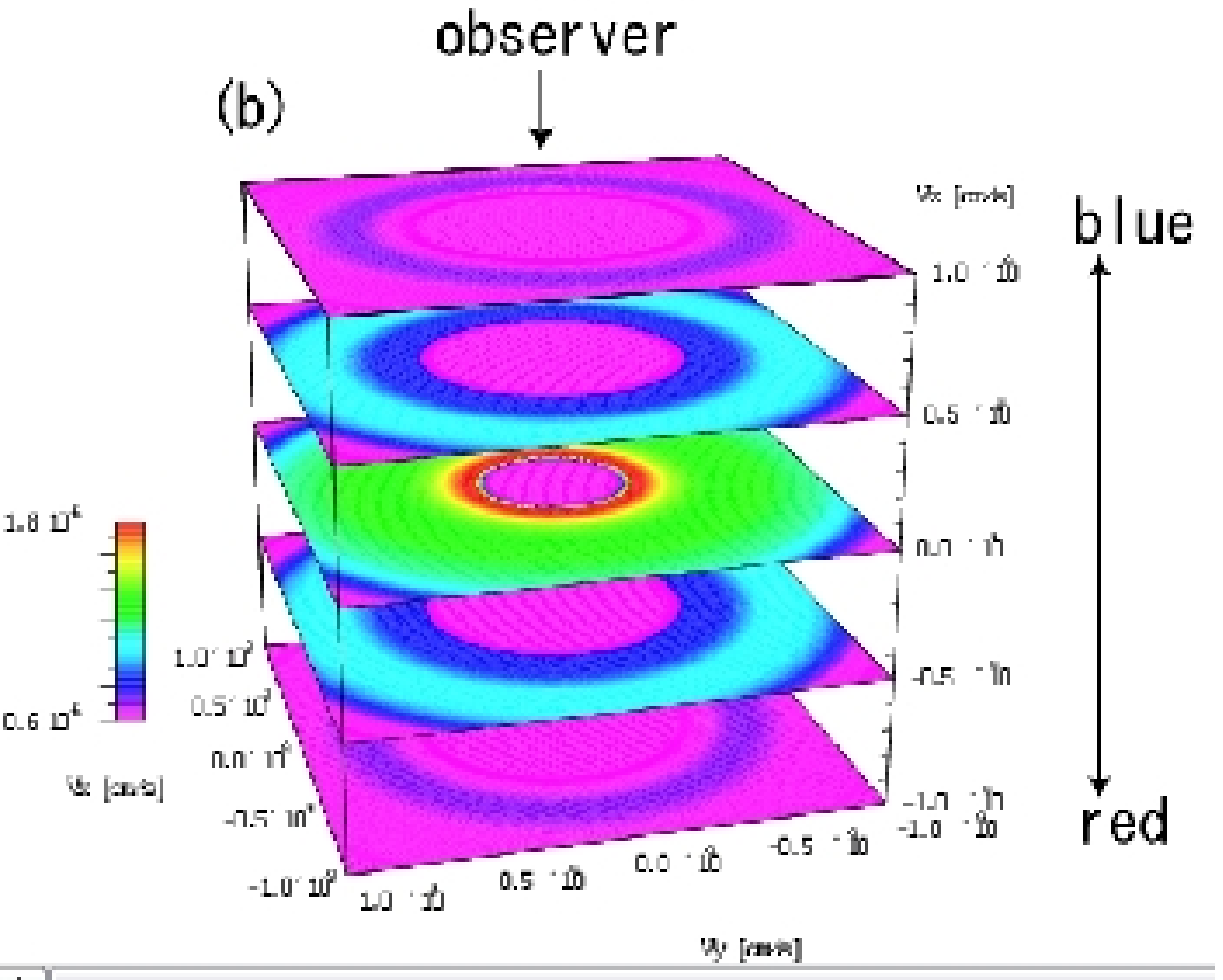,width=1.0\textwidth}
	\end{minipage}
	\begin{minipage}[t]{0.45\textwidth}
		\psfig{file=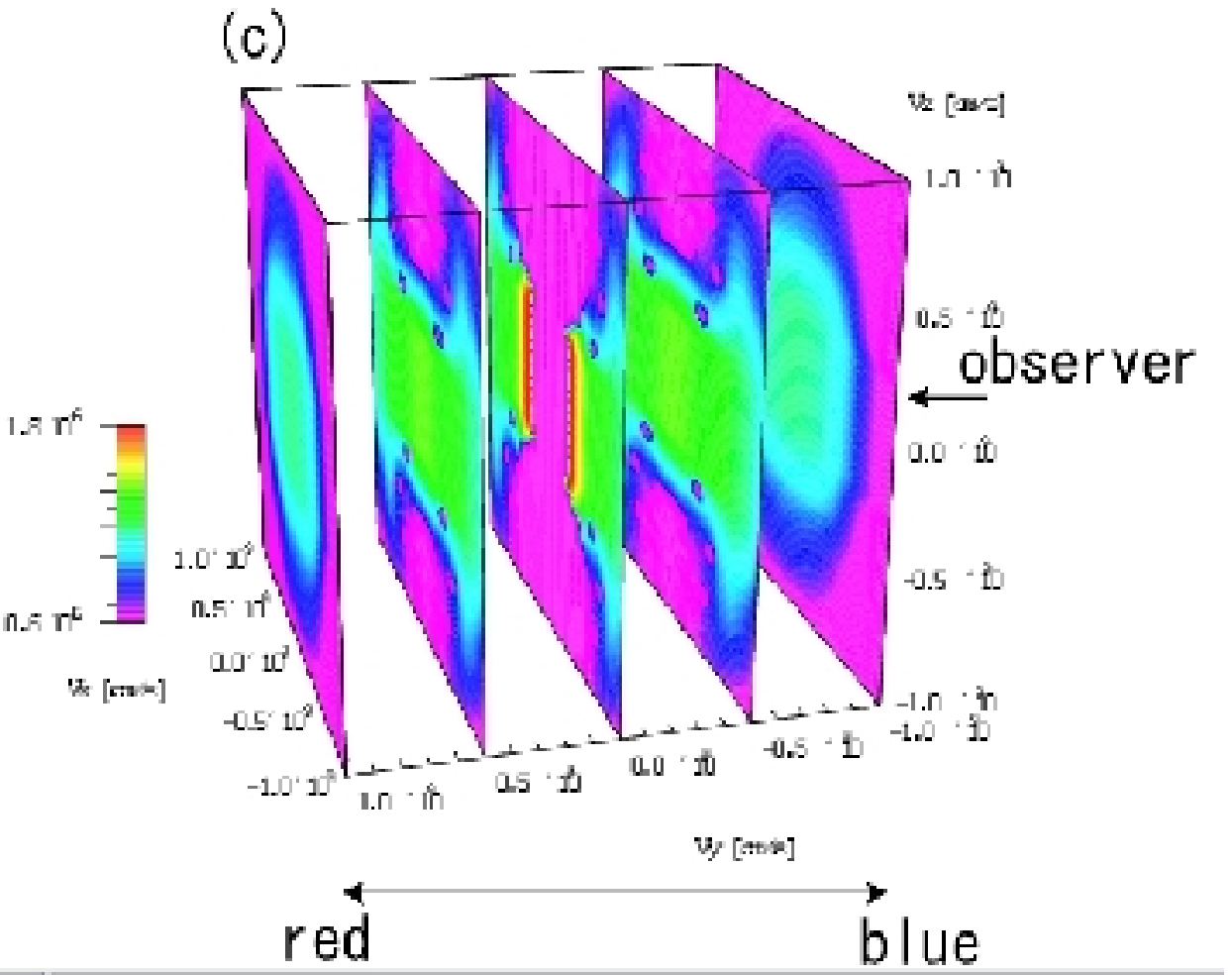,width=1.0\textwidth}
	\end{minipage}\\\hspace{1cm}
       \begin{minipage}[t]{0.3\textwidth}
		\psfig{file=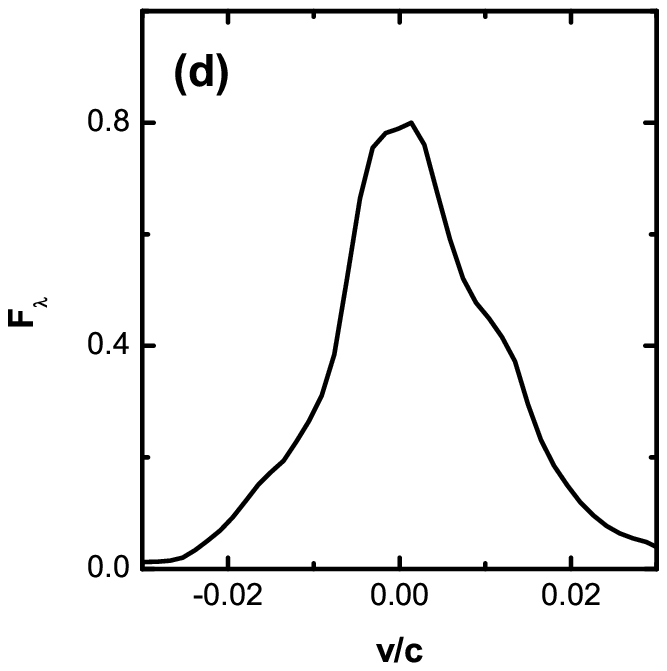,width=1.0\textwidth}
	\end{minipage}\hspace{3cm}
	\begin{minipage}[t]{0.3\textwidth}
		\psfig{file=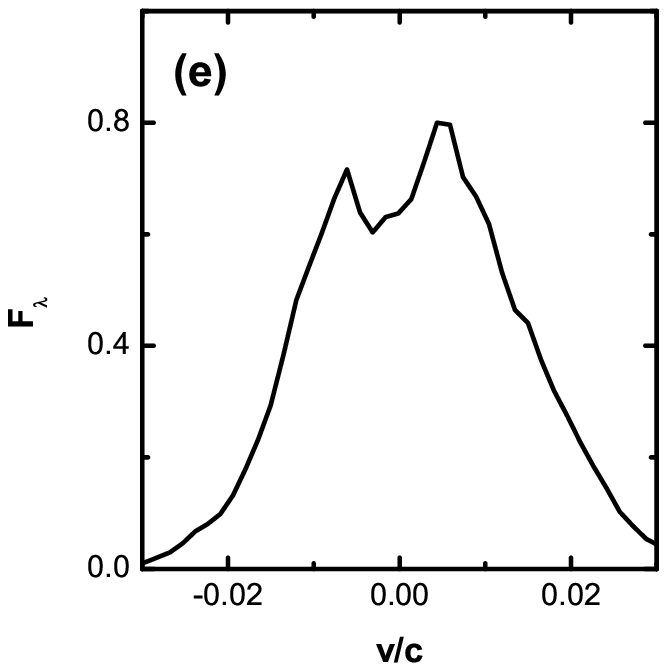,width=1.0\textwidth}
	\end{minipage}\\
\noindent Figure S1: 
The bipolar Model BP8 ({\it S23, S24}). 
(a) Distribution of $^{56}$Ni (which decays into cobalt and then iron, shown in blue) 
and oxygen (red). Density contours are shown covering two orders of magnitude 
and divided into 10 equal intervals on a logarithmic scale. 
(b) The ``SN scan'' for the [\OI] line as seen from an observer placed 
near the axis of the jet ($z$). 
(c) The SN scan for [\OI] for an observer in the equatorial direction 
($x-y$ plane). 
(d) The theoretical [\OI] profile for an observer along the $z$-direction, 
and (e) along the equatorial direction. 
\end{figure*}

\clearpage
\begin{figure*}
\psfig{file=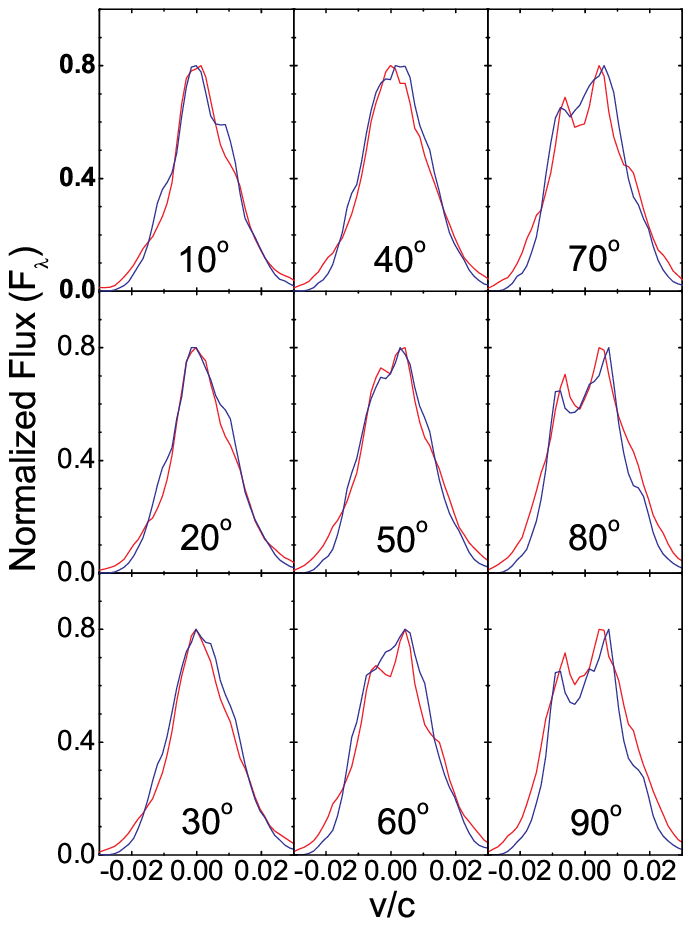,width=0.8\textwidth}
\noindent Figure S2: 
Predicted [\OI] line profiles 
in the aspherical, jet-driven models BP8 (red) and BP2 (blue) 
as a function of the viewing angle $\theta$ ({\it S24}).
\end{figure*}

\clearpage
\begin{table*}
\begin{center}
\noindent Table S1: Supernova Samples
\end{center}
\label{fig:appendix1}
\begin{center}
\begin{tabular}{llllll}\hline
SN & Type & Observing Date & Days\footnotemark[1] & References & Figure\footnotemark[2] \\\hline
1991N  & Ic & 1992-1-9 & 286 & Lick ({\it S1}) & $\circ$ \\\hline
1997dq & broad-Ic & 1998-8-18 & 289 & Lick ({\it S1}) & $\circ$ \\\hline
1997ef & broad-Ic & 1998-9-21 & 299 & Lick ({\it S1}) & $\circ$ \\\hline
2002ap & broad-Ic & 2002-9-15 & 229 & Subaru ({\it S12}) & $\circ$ \\\hline
2003jd & broad-Ic & 2004-9-12 & 323 & Subaru ({\it S11}) & $\circ$ \\\hline
2004dk & Ib & 2005-7-6   & 342 & Subaru & \\
       & & 2005-8-26  & 392 & Subaru & $\circ$ \\\hline
2004fe & Ic & 2005-7-6   & 250 & Subaru& $\circ$ \\
       & & 2005-8-26  & 300 & Subaru & \\\hline
2004gk & Ic & 2005-7-6   & 224 & Subaru & $\circ$ \\\hline
2004gn & Ic & 2005-7-6   & 218 & Subaru & $\circ$ \\\hline
2004gq & Ib & 2005-8-26  & 258 & Subaru & \\
       & & 2005-10-25 & 318 & Subaru & $\circ$ \\
       & & 2005-12-27 & 381 & Subaru & \\\hline
2004gv & Ib/c & 2005-7-6   & 206 & Subaru & \\
       & & 2005-8-26  & 256 & Subaru & $\circ$ \\\hline
2005aj & Ic & 2005-8-26  & 189 & Subaru & \\
       & & 2005-10-25 & 249 & Subaru & \\
       & & 2005-12-27 & 312 & Subaru& $\circ$ \\\hline
2005kl & Ic & 2006-6-30  & 220 & Subaru & $\circ$ \\
       & & 2006-12-25 & 398 & Subaru & \\\hline
2005kz & broad-Ic & 2006-6-30  & 211 & Subaru & $\circ$ \\\hline
2005nb & broad-Ic & 2006-6-30  & 195 & Subaru & $\circ$ \\
       & & 2007-1-24  & 403 & Subaru & \\\hline
2006F  & Ib & 2006-6-30  & 173 & Subaru & \\
       &    & 2006-11-16 & 312 & VLT & $\circ$ \\\hline
2006T  & IIb & 2006-11-26 & 301 & Subaru & \\
       & & 2006-12-25 & 329 & Subaru & $\circ$ \\
       & & 2007-2-18 & 384 & VLT & \\\hline
2006ck & Ic & 2007-1-24  & 249 & Subaru & $\circ$ \\\hline
\end{tabular}\\
\end{center}
\footnotetext{}{$^{1}$Days since the discovery.\\
$^{2}$The spectra shown in Figures 1 and 2 in the main text are 
marked by circles.}
\end{table*}

\clearpage

\subsection*{References} 

\begin{itemize}

\item[S1.]
T. Matheson, A. V. Filippenko, 
W. Li, D. C. Leonard, \& J. C. Shields, 
{\it Astron. J.} {\bf 121}, 1648 (2001). 

\item[S2.]
T. Nakamura, P. A. Mazzali, K. Nomoto, \& K. Iwamoto,
{\it Astrophys. J.} {\bf 550}, 991 (2001). 

\item[S3.]
Ph. Podsiadlowski, P. A. Mazzali, K. Nomoto, 
D. Lazzati, \& E. Cappellaro, 
{\it Astrophys. J.} {\bf 607}, L17 (2004). 

\item[S4.]
K. Nomoto, et al., 
{\it Il Nouvo Cimento} {\bf 121B}, 1207 (2007) 
(astro-ph/0702472).

\item[S5.]
E. Pian, et al.,
{\it Nature} {\bf 442}, 1011 (2006). 

\item[S6.]
F. Patat, et al., 
{\it Astrophys. J.} {\bf 555}, 900 (2001). 

\item[S7.]
G. Kosugi, et al., 
{\it Publ. Astron. Soc. Japan} {\bf 56}, 61 (2004). 

\item[S8.]
K. Maeda, et al.,
{\it Astrophys. J.} {\bf 666}, 1069 (2007). 

\item[S9.]
K. Maeda, et al., 
{\it Astrophys. J.} {\bf 658}, L5 (2007).

\item[S10.]
P. A. Mazzali, et al., 
{\it Astrophys. J.} {\bf 661}, 892 (2007). 

\item[S11.]
P. A. Mazzali, et al., 
{\it Science} {\bf 308}, 1284 (2005). 

\item[S12.]
P. A. Mazzali, et al., 
{\it Astrophys. J.} 
{\bf 670}, 592 (2007). 

\item[S13.]
K. Maeda, P.A. Mazzali, \& K. Nomoto, 
{\it Astrophys. J.} {\bf 645}, 1331 (2006). 

\item[S14.]
K. Nomoto, et al.,
{\it Nature} {\bf 364}, 507 (1993). 

\item[S15.]
D. N. Sauer, et al.,
{\it Mon. Not. R. Astron. Soc.} {\bf 369}, 1939 (2006). 

\item[S16.]
P. A. Mazzali, K. Iwamoto, \& K. Nomoto, 
{\it Astrophys. J.} {\bf 545}, 407 (2001).

\item[S17.]
P. A. Mazzali, et al.,
{\it Astrophys. J.} {\bf 572}, L61 (2002). 

\item[S18.]
P. A. Mazzali, et al.,
{\it Astrophys. J.} {\bf 599}, L95 (2003). 

\item[S19.]
S. Valenti, et al.,
{\it MNRAS} {\bf in press} (2008). 

\item[S20.]
P. A. Mazzali, et al., 
{\it Astrophys. J.} {\bf 645}, 1323 (2006). 

\item[S21.]
N. Tominaga et al.,
{\it Astrophys. J.} {\bf 633}, L97 (2005). 

\item[S22.]
P. A. Mazzali, et al., 
{\it Nature} {\bf 442}, 1018 (2006). 

\item[S23.]
K. Maeda, et al., 
{\it Astrophys. J.} {\bf 565}, 405 (2002). 

\item[S24.]
K. Maeda, K. Nomoto, P. A. Mazzali, \& J. Deng, 
{\it Astrophys. J.} {\bf 640}, 854 (2006).

\item[S25.]
H. Li \& R. McCray, 
{\it Astrophys. J.} {\bf 387}, 309 (1992). 

\end{itemize}

\end{document}